\documentstyle{article}
\begin{document}
\author{O.D.Dalkarov}
\title{On the nuclear reactions with participation of nearthreshold baryon resonances}
\date{Lebedev Physical Institute, Moscow, Russia}
\maketitle

\begin{abstract}
     The main features of nuclear reactions with participation of
baryon resonances ($\Delta$(1232), S(1535) and $\Lambda$(1405)),
which are interpreted as a manifestation of the bound states of
nucleus and corresponding meson are discussed.
\end{abstract}

\section{Pion-nuclear interaction in resonance energy region}

   In the first experimental works for $\pi$-meson interaction with
nuclei in the energy diapason of the order of 100 MeV and higher
the specific behaviour caused by the resonance nature of
$\pi$-meson interaction with nuclear nucleon or with nuclei as
whole were revealed. From the experimental point of view the
reactions with simplest nucleus - deutron, in particular, $\pi^+$+
d$\to$p + p, $\pi^+$ + d $\to$ $\pi^+$ + p + n, $\pi$d-elastic
scattering and others were most investigated. The reactions
enumerated above are followed to common peculiarity: the behaviour
of total cross-sections and others characteristics are
corresponded to resonanced nature of $\pi$d - interaction at the
energy about 150 MeV ( in the c.m.s. ). In particular, first
experimental data obtained by M.G.Mescheryakov et al.~\cite{1}
about behaviour of total cross - sections of $\pi^+$d$\to$pp
reaction and inverse one show distinquishable maximum in this
energy region ( with the width of the order of 120 MeV ). The
analysis of angular distribution of $\pi^+$-meson for p + p$\to$ d
+ $\pi^+$ reaction points to P-wave meson production and besides
that the angular distribution is the same with the increasing of
$\pi$-meson energy by the factor two. This fact can indicate
either on the equal ratio between $^{1}S_{0}\to^{3}S_{1}$ and
$^{1}D_{2}\to^{3}S_{1}$ transitions to whole considered energy
diapason and amplitudes for these transitions should be equally
depend on the energy or the dominant role plays one of these
transitions.

   Experimental data could be qualitativaly explained if one take
into account that the $\pi$-meson interaction has a resonance
behaviour in the state with spin and isospin are equal 3/2 (isobar
$\Delta_{33}$) In this case the main contribution to P-wave
$\pi$-meson production will be caused by $^{1}D_{2}\to^{3}S_{1}$
transition. There are two possibilities for explanation of
resonance maximum in the reactions $pp\to d\pi^{+}$ and
$\pi^{+}d\to pp$ : either two-nucleon interaction
in$^{1}D_{2}$-state has resonance nature, i.e. so called
"dinucleon resonance" is formed (in other terminology - simplest
"isonucleus" or "$\pi$- mesonucleus") or this process takes place
throw the virtual production of $\Delta_{33}$-isobar.

    The theoretical consideration of these reactions in the
assumption of virtual isobar-nucleon state was carried out in
~\cite{2}, where triangle Feynman diagram with virtual isobar was
calculated and it was shown that the energetical behaviour near
complex singularity of considered diagram could immitate resonance
behaviour of total cross - section for $\pi^{+}d\to pp$ reaction.
In this case angular distribution of $\pi$-meson is caused by
dominant contribution of S-wave in the relative motion of
$\Delta_{33}$ and nucleon. By the analogy resonance behaviour of
$\pi^{+}d\to pp$ reaction in the energy region near 2.84 GeV could
be explained by the virtual production of $\Delta(1920)$~\cite{3}.

    As for the experimental investigation of $\pi$d elastic scattering
in resonance region i.e. at the $\pi$-meson energies of the order
of 200 MeV( in the l.s. ) it was also revealed the existence of
distinquishable maximum in the total cross - section. Observed
maximum in the total cross - section and ratio
$\left(\sigma(\pi^{-}d)-\sigma(\pi{-}p)\right)/\sigma(\pi^{-}p)\approx
3$ indicate on virtual $\Delta_{33}$-nucleon production.

    On the language of Feynman diagrams this mechanism is
corresponded to the calculation of nonrelativistic squared
diagram. The total and differential cross - sections of $\pi$d
elastic scattering were calculated with this diagram ~\cite{4}.
Resonance behaviour is caused by singularities of this diagram
which lie near physical energy region. The comparison with
experimental data show rather well aggreement ( some differences
between the calculations correspond to the different deutron
form-factors at the relatively large ($ >$ 150 MeV/c ) relative
momenta). Precision investigation of differential cross - section
could be used as a criterium between two possible explanations for
observed resonance maximum: a) formation of two - nucleon
resonance ( "isonucleus" or "mesonucleus" ) with B = 2 and b)
enchancement mechanism due to singularities of Feynman diagrams
just near physical region of kinematical variables.

    Thus the exact conclusion about the existence of simplest
"isonucleus" ( or "mesonucleus" ) is not possible to make from the
contemporary experimental data and theoretical works since it is
necessary to exclude the contribution of peripheral mechanisms
which are corresponded to $\Delta_{33}$-nucleon virtual state.

    As for the interaction of $\pi$-meson with complex nuclei the
interesting experimental observation was made for the reaction
$\pi^{-}{^{12}C} \to \pi^{-}n^{11}C$ where the resonance maximum
in the total cross - section at the kinetic $\pi$-meson energy of
the order of 200 MeV ( in the l.s. ) was observed ~\cite{5}. Using
the well known theory of direct nuclear reactions the analysis of
this reaction was made~\cite{6}. The behaviour of total
cross-section at $\pi$-meson energies more than 200 MeV was well
quantitatively described by the single nucleon mechanism ( pole
approximation ) for the production of $\Delta_{33}$ and its
following decay to $\pi$-nucleon channel. However in the low
energy region ( $< 200$ MeV ) there is significant discrepance
between experimental data and calculations with pole
approximation. Finally, maximum in the total cross-section for the
$^{12}C{\pi^{-} \to \pi^{-}n}^{11}C$ reaction was found to be
wider and shifted in comparison with pole approximation.

    The reason for this discrepance (as it was shown~\cite{7}) could
be consist in the assumption that not only pole diagram is
distinquished in this energy region. Namely, one can be noted that
the triangle diagram, which corresponds to elastic scattering of
$\Delta_{33}$ on $^{11}C$, has the singularity just near physical
region of kinematical variables. In this case we deal with the
complex singularity at the energy ( in the c.m.s. ) is equal to
the sum of the masses of $\Delta_{33}$ and $^{11}C$ nucleus.
Taking into account the contribution of pole and triangle diagrams
the rather well description of the experimental data was obtained.
In particular, the broadering of resonance maximum and its shift
(of the order of 30 MeV ) to the region of smaller energy of
incident $\pi$-mesons were explained. Note, that the contribution
of diagram with rescattering of $\Delta$-isobar on intermediate
nucleus looks like as interaction of "stopped" isobar with
$^{11}C$ nucleus.

     Thus the present experimental data and theoretical
considerations leave the question open on the existence of
"isonuclei". In each concrete case it is necessary to carry out
the complex of experimental and theoretical investigations to
separate the effects of nuclear reaction mechanism from the
formation of "isonuclei" i.e. bound state of isobar ( or -
$\pi$-meson ) with nucleus. First of all it is needed to
investigate the angular distributions and polarized
characteristics.

\section{ $\eta$-nuclei}

    At present there are considerable experimental data and
theoretical investigations on the possible existence of
$\eta$-meson-nucleus bound states or "$\eta$-nuclei". Detailed
review of present situation was done in~\cite{8}. Difficulties in
interpretation of experimental data consist in the production type
mechanism for $\eta$-meson on free nucleon which has threshold
character and is determined by the production of $S_{11}$(1535)
baryon resonance and its following decay to $\pi$N or $\eta$N with
equal probabilities. Most convinced data on the photoproduction of
$S_{11}$(1535) on nuclei were obtained using the beam of
$\gamma$-quanta from electron sinchrotron (FIAN).The spectrum of
correlated $\pi^{+}n$ pairs arising from
$\gamma+^{12}C\to\pi^{+}nX$ and flying transversely to photon beam
have been observed~\cite{9}. The main result consists in
observation of the shift for resonance maximum in ($\pi^{+}n$)
system in comparison with Table value for $S_{11}$(1535).

     In view of these experimental data one could takes notice of
the following: a. observed shift in the position of $S_{11}$(1535)
in the reaction $\pi^{+}+^{12}C\to \pi^{+}nX$ looks like as the
same phenomenon for production of $\Delta_{33}$(1232) in the
reaction $\pi+^{12}C\to \pi$NX. In the last case observed shift
could be explained by the possible nuclear reaction mechanism for
production of $\Delta_{33}$ on nucleus. This mechanism includes
triangle Feynman diagram with rescattering $\Delta_{33}$ on
intermediate nucleus besides pole diagram corresponded to the
production on the bound nucleon. Taking into account these two
diagrams it was turn out to explain the broadering and shift of
$\Delta_{33}$ in production on nuclei. Note that in rescattering
processes as a propagater for virtual $\Delta$ it is necessary to
use a pole position in the $\pi$N amplitude which corresponds to
smaller mass in comparison with the value on mass shell ( Re
M($\Delta_{33}$)= 1210 MeV for $\Delta_{33}$(1232)). The same pole
position for $S_{11}$(1535) is equal to Re M($S_{11}$) = 1500 MeV.
At the same time the position of complex singularity is determined
by the sum of Re M(resonance) and mass of nucleus. To clear up a
nuclear reaction mechanism for photoproduction of $S_{11}$(1535)
on nuclei the additional experimental and theoretical efforts are
needed. As for the one nucleon mechanism (pole diagram ) this
question was considered in details in~\cite{10}. In particular an
investigation of small transvers momentum region for ($\pi^{+}n$)
pairs gives possibility to use the Treyman - Yang
criterium~\cite{11}.

     In view of considered above the investigation of peripheral
mechanisms for production of $S_{11}$(1535) on nuclei can give
additional information about the phenomena connected with the
formation of $\eta$-nuclei on the background of peculiarities of
the nuclear reaction mechanism.

\section{ K - nuclei }

    By the analogy with the cases considered above the existence of
nearthreshold $\Lambda$(1405) baryon resonance determining low
energy KN - interaction as input for hypothesis about possible
formation of nearthreshold K - nucleus bound states was used.

    In the work~\cite{12} phenomenological potential model for
K-meson interaction with light nuclei was proposed. For elementary
KN-interaction a potential approximation with simplest Gaussion
form for interaction potential was used. The parameters of KN -
potential were choosed so to obtain correct KN scattering lengths
known from the shifts and widths of $K^{-}p$ and $K^{-}d$ atoms
and to reproduce the mass of $\Lambda$(1405) as a bound state of
K-meson and nucleon. Using obtained so two-body KN - potential the
optical potential for K - nucleus interaction was calculated. With
this optical potential the existence of $^{3}_{K}H$(I=0,1) and
$^{4}_{K}H$(I=1/2) bound states with binding energies from 20 MeV
up to 100 MeV and widths of the order several tens MeV were
predicted.

     The proton and neutron energy distributions from He(stopped $K^{-}{,}N$)
reactions were measured in~\cite{13}. In the missing mass spectrum
two mono-energetic peaks were observed. These peaks peaks were
interpreted as the formation of tribaryon $S^{0}$(3115) and
$S^{+}$(3140) with isospins I=0 and 1 and strangeness S=-1,
correspondingly. In the potential model considered in~\cite{12}
these states should correspond to the bound $^{3}_{K}H$(I=0) and
$^{3}_{K}H$(I=1) states. Binding energies for these states ($~$200
MeV) turned out to be larger and widths smaller ( $<$20 MeV ) than
theoretically predicted. Nevertheless taking into account
statistical significance (~13$\sigma$ for S(3115) and ~3.7$\sigma$
for S(3140)) these experimental results really indicate on the
possible existence of three-body K-nuclei. Another interpretations
of these data are not so convincing ( for instance, the
interpretation as the $\Sigma$NN hypernuclei is less probable
since excitation energies for these hupernuclei should be of the
order 50 MeV or higher ).

     As the direct test to prove the existence of bound K-nuclei it
will be useful to observe the discrete spectra of $\gamma$-quanta
or $\pi$-mesons injected due to transitions from states of
$K^{-}$-He atom to one of the states of three-body K-nucleus.
Simple estimations give for the probabilities of radiative
transitions the value of the order of $10^{-3}-10^{-4}$ and in the
case of probabilities for $\pi$-meson transitions will be
$10^{-1}-10^{-2}$, correspondingly.

      In conclusion, it would be note that the question on the
existence of new kind of nuclei with meson as a constituent is a
great of interest. However the interpretation of present
experimental data is not so simple. From our point of view the
reasons are either in the necessity of experimental confirmation
for obtained data ( for instance, in the case of K -nuclei ) or in
the additional experimental and theoretical analysis. In
particular the question about possible nuclear reaction mechanism
should be especially checked. Here we are not considered the
predictions of quarks model in which the rich spectrum of
multiquark states are predicted~\cite{14}since to speak about
quark nature of the states considered above without identification
even if one of ground states looks like as prematurely.

   This work is supported by the RFBR grant 05-02-17482-a.

\end{document}